\documentclass[onecolumn, draft, 12pt]{IEEEtran} 
\usepackage{amsmath,cases}
\usepackage{amsthm}
\usepackage{amssymb}
\usepackage{cite}
\usepackage{amsfonts}
\usepackage{enumerate}
\usepackage{url}
\usepackage[final]{graphicx}
\usepackage[mathscr]{euscript}
\newtheorem{lem}{Lemma}
\newtheorem{defn}{Definition}

\newtheorem{proposition}{Proposition}

\newcommand{\vect}[1]{\ensuremath{\mathbf{\lowercase{#1}}}} 
\newcommand{\mat}[1]{\ensuremath{\mathbf{\uppercase{#1}}}} 
\newcommand{\set}[1]{\ensuremath{\mathscr{\uppercase{#1}}}} 
\newcommand{\PSD}[1]{\mathbb{S}^{#1}_{+}}
\newcommand{\eig}[2]{\ensuremath{\lambda_{#1}(\mat{#2})}}
\newcommand{\re}{\mathbb{R}}
\newcommand{\N}{\mathbb{N}}
\newcommand{\vectornorm}[1]{\left|\left|#1\right|\right|}
\newcommand{\E}[1]{\ensuremath{\mathbb{E}\left[ #1 \right]}}
\newcommand{\Tr}[1]{{\mathrm{tr} \;}  #1  }
\newcommand{\Det}[1]{{\mathrm{det}} \left( #1  \right)}
\newcommand{\mlog}{\mathbf{\rm log}}
\newcommand{\diag}[1]{\ensuremath{\mathrm{\mathbf{diag}}\left( #1 \right)}}
\newcommand{\card}[1]{\ensuremath{\mathrm{card} \; #1}}
\newcommand{\pdf}[2]{\ensuremath{f_{#1}\left(#2 \right)}}

\newcommand{\BF}{\ensuremath{\mathcal{BF}}}
\newcommand{\TBF}{\ensuremath{\mathcal{TBF}}}
\newcommand{\TBFm}{\ensuremath{\mathcal{TBF}_{m}}}
\newcommand{\cTBF}{\ensuremath{\mathcal{CTBF}}}
\newcommand{\cTBFm}{\ensuremath{\mathcal{CTBF}_{m}}}
\newcommand{\cmimomac}[1]{\ensuremath{\set{C}^{#1}_{{\rm MIMO-MAC}}}}
\newcommand{\cm}{c.m.}
\newcommand{\TP}[1]{\ensuremath{\mathcal{TP}_{r}}}
\newcommand{\STP}[1]{\ensuremath{\mathcal{STP}_{r}}}
\newcommand{\C}[1]{\ensuremath{\log\left(1+ #1 \right)}}
\newcommand{\LD}[1]{\ensuremath{\log \Det{\mat{I}+ #1}}}

\newcommand{\F}{\ensuremath{\mathcal{G}}}
\newcommand{\orderl}[1]{\ensuremath{\leq_{{\rm #1}}}}
\newcommand{\orderm}[1]{\ensuremath{\preceq_{{\rm #1}}}}

\newcommand{\norderl}[1]{\ensuremath{\nleq_{{\rm #1}}}}
\newcommand{\CDF}[2]{\ensuremath{F_{#1}\left( #2 \right)}}
\newcommand{\PDF}[2]{\ensuremath{f_{#1}\left( #2 \right)}}

\newcommand{\Cerg}[2]{\ensuremath{\overline{C}^{(#1)}_{{\rm #2}}\left( \rho \right)}}
\newcommand{\D}{\ensuremath{\mathtt{d}}}

\begin{document}

\title{Stochastic Ordering of Fading Channels Through the Shannon Transform}
\author{Adithya Rajan, Cihan Tepedelenlio\u{g}lu, \emph{Member, IEEE} \thanks{The authors are with the School of Electrical, Computer, and Energy Engineering, Arizona
State University, Tempe, AZ 85287, USA. (Email:
\{arajan2,cihan\}@asu.edu). This work was supported in part by the National Science Foundation under Grant CCF 1117041. Part of this work has appeared in \cite{paper:AdithyaCapacity12}. } }
\maketitle

\begin{abstract}
A new stochastic order between two fading distributions is introduced. A fading channel dominates another in the ergodic capacity ordering sense, if the Shannon transform of the first is greater than that of the second at all values of average signal to noise ratio. It is shown that some parametric fading models such as the Nakagami-$m$, Rician, and Hoyt are distributions that are monotonic in their line of sight parameters with respect to the ergodic capacity order. Some operations under which the ergodic capacity order is preserved are also discussed. Through these properties of the ergodic capacity order, it is possible to compare under two different fading scenarios, the ergodic capacity of a composite system involving multiple fading links with coding/decoding capabilities only at the transmitter/receiver. Such comparisons can be made even in cases when a closed form expression for the ergodic capacity of the composite system is not analytically tractable. Applications to multiple access channels, and extensions to multiple-input multiple-output (MIMO) systems are also discussed.

\end{abstract}

\IEEEpeerreviewmaketitle

 \section{Introduction}
Consider a flat fading channel with additive white Gaussian noise (AWGN), where the receiver has perfect channel state information (CSI). The maximum achievable rate of this system, when coding is applied across multiple independent channel realizations is known as the ergodic capacity, and is given by $\E{\C{\rho X}}$, where $\rho \geq 0$ represents the average signal to noise power ratio (SNR) of the system, and $\rho X$ represents the instantaneous SNR random variable (RV). This expectation is also known as the Shannon transform of $X$ \cite[pp. 44]{book:verdu}, \cite{paper:letzepis07}.

In this work, a stochastic order which can be used to compare fading channels based on the Shannon transform of the instantaneous SNR is discussed. A fading channel is said to be better than another in the ergodic capacity order, if its corresponding ergodic capacity is bigger for all $\rho$. The proposed order is a kind of stochastic order on positive RVs. Stochastic orders in general find applications in economics \cite{quirk62}, reliability analysis \cite{belzunce04}, and actuarial sciences \cite{book:muller02}. A comprehensive exposition of stochastic orders can be found in \cite{book:shaked}. Previously, the stochastic Laplace transform (LT) order, which compares the real-valued Laplace transforms of RVs has been used to compare two fading distributions and applied to comparing the average error rate of $M$-ary quadrature amplitude modulation ($M$-QAM) \cite{paper:adithya11}. This can be explained by the fact that error rates of some modulations are non negative integral mixtures of decaying exponentials, which can also be viewed as the Laplace transform. It has been shown in \cite{paper:adithya11} that Laplace transform ordering of instantaneous SNRs implies ordering of ergodic capacities, but not conversely. 

The ergodic capacity order presented in Section \ref{J_CapOrd_sec:erg_cap_ord} of this paper is new to both stochastic ordering literature as well as information theory literature. Although this stochastic order was first introduced in \cite{paper:AdithyaCapacity12}, the current paper offers a detailed discussion of its properties, examples and extensions relevant to wireless communications, including the MIMO case. Further, some of the convergence properties of the Shannon transform are also studied. In this paper, many parametric fading distribution families such as the Nakagami-$m$, Rician and Hoyt are observed to have the property that the ergodic capacity is monotone with respect to the line of sight (LoS) parameter for each of these distributions. Consequently, the instantaneous SNR of these fading channels serve as examples of ergodic capacity ordered random variables. The properties of this stochastic order are useful in obtaining comparisons of the performance of systems involving multiple SNR RVs, as described in Section \ref{J_CapOrd_sec:sys_mult_rv}. For example, let $\lbrace X_{i} \rbrace_{i=1}^{M}$ and $\lbrace Y_{i} \rbrace_{i=1}^{M}$ be two sets of fading channels such that the ergodic capacity over $X_{i}$ is less than that of $Y_{i}$, $i=1,\ldots,M$ at all SNR. Then, the properties of the ergodic capacity order provide the conditions under which a composite system consisting of $\lbrace X_{i} \rbrace_{i=1}^{M}$ as the component fading channels has a smaller ergodic capacity than that of a system with components $\lbrace Y_{i} \rbrace_{i=1}^{M}$. Such comparisons of ergodic capacities can be made even in cases when a closed-form expression is not available, such as diversity combining schemes and fading multiple access channels (MAC). A MIMO extension of the definition of the ergodic capacity order, which can be used to order positive semidefinite random matrices is given in Section \ref{J_CapOrd_sec:MIMO}. 

\subsection{Notations and Conventions}
The set of real numbers, positive integers and complex positive semidefinite symmetric matrices of size $n \times n$ are denoted by $\re$, $\N$, and $\PSD{n}$ respectively, while all other sets are denoted using script font. For a finite set $\set{B}$ the cardinality is denoted by $\card{\set{B}}$, while the indicator function is defined as $I(x \in \set{K}) = 1$, if $x \in \set{K}$ and $0$, otherwise. For any measure $\mu(\cdot)$, $\mu(u)$ is used to represent $\mu([0,u])$. Vectors and matrices are denoted by boldface lower-case and upper-case letters respectively. For both the cases, $\vectornorm{\cdot}$ denotes the $L^{2}$ norm. The trace and determinant of a matrix $\mat{M}$ are denoted by $\Tr{\mat{M}}$ and $\Det{\mat{M}}$ respectively. The identity matrix is denoted by $\mathbf{I}$. If $a_{i} \in \re$, $i= 1,\ldots,N$, then $\diag{a_{1},\ldots,a_{N}}$ is the diagonal matrix whose $(i,i)$ element is $a_{i}, i=1,\ldots,N$. The $i^{th}$ smallest eigenvalue of $\mat{A} \in \re^{N \times N}$ is denoted by $\eig{i}{A}$, $i=1,\ldots,N$, and the set of all eigenvalues is denoted by $\mathbf{\Lambda}(\mat{A})$. For a random variable $X$, $\CDF{X}{x}$ and $\PDF{X}{x}$ denote the cumulative distribution function (CDF) and the probability density function (PDF) respectively. $\E{g(X)}$ is used to denote the expectation of the function $g(\cdot)$ over the PDF of $X$. All logarithms are natural logarithms. We write $f_{1}(x) = O(f_{2}(x))$, $x \rightarrow a$ to indicate that $\limsup_{x \rightarrow a} (f_{1}(x)/f_{2}(x)) < \infty$. 

\section{Mathematical Preliminaries}

\subsection{Completely Monotone Functions}
\label{J_CapOrd_subsec:math_cm}
A function $g : (0,\infty) \rightarrow \re$ is said to be completely monotone (c.m.), if it possesses derivatives of all orders which satisfy
\begin{align}
\label{J_CapOrd_eqn:cm_def}
(-1)^{n} \frac{\D ^{n}}{\D x^{n}} g(x) \geq 0 ,
\end{align}
for all $x \geq 0$ and $n \in \N \cup \lbrace 0 \rbrace$, where the derivative of order $n=0$ is defined as $g(x)$ itself. The celebrated Bernstein's theorem \cite{book:schilling} asserts that, $g : (0,\infty) \rightarrow \re$ is c.m. if and only if it can be written as a mixture of decaying exponentials: 
\begin{equation}
\label{J_CapOrd_eqn:bernstein}
g(x) = \int\limits_{[0,\infty)} \exp(-u x)  \mu(\D u) ,
\end{equation}
which is a Lebesgue integral with respect to a positive measure $\mu$ on $[0,\infty)$. By definition, c.m. functions are positive, decreasing and convex, and it is straightforward to verify that positive linear combinations of c.m. functions are also c.m. \cite{book:schilling}.

\subsection{Stieltjes Functions}
The set of Stieltjes functions is a subclass of the set of completely monotone functions, and is denoted by $\set{S}$. A function $g: (0,\infty) \rightarrow [0,\infty)$ is said to belong to $\set{S}$ if it admits the representation
\begin{align}
\label{J_CapOrd_eqn:st_nec_suf}
g(x) = a/x + b + \int\limits_{(0,\infty)} (x+u)^{-1}\mu(\D u) \;,
\end{align}
where $a,b \geq 0$, and $\mu$ is a nonnegative measure on $(0,\infty)$ which satisfies the convergence condition $\int_{(0,\infty)}(1+u)^{-1}\mu( \D u) < \infty$. It is easy to show that any Stieltjes function is also a double Laplace transform of a nonnegative function. A necessary and sufficient condition for $x \mapsto g(x) \in \set{S}$ is that $x \mapsto (g(x^{-1}))^{-1}$ also belongs to $\set{S}$ \cite[p. 66]{book:schilling}.
\subsection{Bernstein Functions}
A function $g: (0,\infty) \rightarrow \re$ is a Bernstein function, if $g(x) \geq 0, \forall x >0$, and $\D g(x)/\D x$ is {\cm} Equivalently, $g(x)$ admits the representation \cite[p. 15]{book:schilling}
\begin{align}
\label{J_CapOrd_eqn:bernst_nec_suf}
g(x) = a+bx+\int\limits_{(0,\infty)} \left( 1 - \exp(-ux)\right) \mu(\D u) \;,
\end{align}
for some $a,b \geq 0$, where $\mu$ is a nonnegative measure on $(0,\infty)$ satisfying $\int_{(0,1)} \mu(\D u)  + \int_{[1,\infty)}u \mu(\D u) < \infty$. The set of all Bernstein functions is denoted by {\BF}. 

An important property is that the set {\BF} is closed under positive linear combinations: if $g_{i} \in$ {\BF}, and $a_{i} \geq 0$, $i = 1,\ldots,N$, then $\sum_{i=1}^{N}a_{i}g_{i} \in$ {\BF}. Some examples of Bernstein functions are $g(x) = x^{\alpha}$, for $0 < \alpha <1$, $g(x) = x/(1+x)$ and $g(x) = \log(1+x)$. The representation of the capacity function $\log(1+x)$ in the form \eqref{J_CapOrd_eqn:bernst_nec_suf} is known as Frullani's integral \cite[p. 6]{book:lebedev72}, and is given by 
\begin{align}
\label{J_CapOrd_eqn:cap_alt}
\C{x} = \int\limits_{0}^{\infty} \left( 1 - e^{-sx}\right) \frac{e^{-s}}{s} \D s \;.
\end{align}

\subsection{Thorin-Bernstein Functions}
A Bernstein function $g$ is called a Thorin-Bernstein function \cite[pp. 73-79]{book:schilling}, if it admits the representation given by \eqref{J_CapOrd_eqn:bernst_nec_suf}, where $s\mu(s)$ is {\cm} The family of all Thorin-Bernstein functions is denoted by {\TBF}. A necessary and sufficient condition for $g: (0,\infty) \rightarrow (0,\infty)$ to be in {\TBF} is that $g$ can be represented as follows \cite[p. 73]{book:schilling}:
\begin{align}
\label{J_CapOrd_eqn:tbf_nec_suf}
g(x) = a+bx+\int\limits_{(0,\infty)} \log \left( 1 + x/s\right) \mu(\D s) \;,
\end{align}
for some $a,b \geq 0$ and $\mu$ is a positive measure on $(0,\infty)$, which satisfies the convergence condition $\int_{0}^{1}|\log s|\mu(\D s)  + \int_{1}^{\infty}s^{-1}\mu(\D s) < \infty$. We refer to any $g_{2} \in \TBF$ which satisfies the property that $g_{1}(g_{2}(\cdot)) \in \TBF$ for all $g_{1} \in \TBF$ as a \emph{composable} Thorin-Bernstein function (we denote the set of all such functions by {\cTBF}). A necessary and sufficient condition for any $g_{2}$ to belong to $\cTBF$ is that $(\D g_{2}(x)/\D x)/g_{2}(x) \in \set{S}$ \cite[Theorem 8.4]{book:schilling}. Functions belonging to the class {\TBF} are of particular relevance to this paper, since the Shannon capacity function $C(x) := \C{x}$ not only belongs to {\BF}, but also belongs to {\TBF}, as seen from \eqref{J_CapOrd_eqn:cap_alt} and \eqref{J_CapOrd_eqn:tbf_nec_suf}.

It is useful to define a multivariate extension of a Thorin-Bernstein function. A function $g: \re^{m} \rightarrow \re$ belongs to {\TBFm} if $g(x_{1},\ldots,x_{m})$ is a Thorin-Bernstein function in each argument, when all other arguments are treated as constants. Further, if $g$ is composable in each variable when all other variables are fixed, then $g$ is said to belong to the set {\cTBFm}. An example of function in {\cTBFm} can be verified to be $g(x_{1},\ldots,x_{M}) = \sum_{i=1}^{M} \alpha_{i} x_{i}, \alpha_{i} \geq 0, i = 1,\ldots, M$.

\subsection{Matrix Functions}
\label{J_CapOrd_subsec:mat_fun}
Let $\phi: \re \rightarrow \re$, and $\lambda_{i} \in \re$, $i=1,\ldots,N$. If $\mat{D}  = \diag{\lambda_{1},\ldots,\lambda_{N}}$, we define $\phi(\mat{D}) = \diag{\phi(\lambda_{1}),\ldots,\phi(\lambda_{N})}$. If $\mat{A} \in \PSD{n}$, so that $\mat{A} = \mat{U} \diag{\lambda_{1}(\mat{A}),\ldots, \lambda_{N}(\mat{A})} \mat{U}^{\rm H}$, where $\mat{U}$ is a unitary matrix, then we define $\phi(\mat{A}) = \mat{U} \phi(\mat{D}) \mat{U}^{\rm H}$, provided $\phi$ is well defined on the eigenvalues of $\mat{A}$. In this way, $\phi(\mat{A})$ can be defined for all Hermitian matrices of any order \cite{book:bhatia1997}. In this work, the scalar function and its matrix extension are denoted using the same symbol, and the argument of the function defines the specific context. Matrix functions find applications in Section \ref{J_CapOrd_sec:MIMO}. 
We also use multivariate functions with matrix arguments in Section \ref{J_CapOrd_sec:MIMO}, which are defined through the Cauchy integral formula as given in \cite{paper:matrixFunctionPaper}. While we refrain from providing the explicit definition here due to its rather technical nature, it suffices to note that such functions satisfy the following two properties \cite{paper:matrixFunctionPaper}, which will be used in our work. 
\begin{lem}
\label{lem:multivMatrxFuntrace}
If $\mat{A}_{m} \in \PSD{n}$, $m=1,\ldots,M$ then
\begin{align}
\Tr{f(\mat{A}_{1},\ldots,\mat{A}_{M})} = \sum\limits_{i_{1}=1}^{n}\ldots\sum\limits_{i_{M}=1}^{n} f(\lambda_{i_{1}}(\mat{A}_{1}),\ldots,\lambda_{i_{M}}(\mat{A}_{M}))\;.
\end{align}
\end{lem}
\begin{lem}{{\cite[Theorem 3.4]{paper:matrixFunctionPaper}, \cite[p. 13]{paper:matrixFunctionPaper}}}
\label{lem:multivMatrixFunComp}
Let $\mat{A}_{m} \in \PSD{n}$, $m=1,\ldots,M$. If $f$ is a multivariate matrix function well defined on the eigenvalues of $\mat{A}_{m}$, and $\phi$ is a univariate matrix function which is well defined on the eigenvalues of $f(\mat{A}_{1},\ldots,\mat{A}_{m})$, then $\phi (f(\mat{A}_{1},\ldots,\mat{A}_{m})) = (\phi \circ f)(\mat{A}_{1},\ldots,\mat{A}_{m})$.
\end{lem}
\subsection{Integral Stochastic Orders}
Let $\F$ denote a class of real valued functions $g : \re^{+}  \rightarrow \re$, and $X$ and $Y$ be random variables (RVs). We define the integral stochastic order with respect to $\F$ as \cite{book:muller02}:
\begin{align}
\label{J_CapOrd_eqn:integral_st_order_def}   
X \leq_{\F} Y \iff \E{g(X)} \leq \E{g(Y)} \;,\; \forall g \in \F\; .
\end{align}
In this case, $\F$ is known as a generator of the order $\orderl{\F}$. We now give an example of an integral stochastic order relevant to this paper, by specifying the corresponding generator set of functions $\F$.
\subsubsection{Laplace Transform Order}
\label{J_CapOrd_subsec:lt_order}
This partial order compares random variables based on their Laplace transforms. Here, $\F = \lbrace g(x) : g(x) = - \exp\left( - \rho x \right) , \; \rho \geq 0 \rbrace$, so that $X \orderl{Lt} Y$ is defined as 
\begin{align}
\E{\exp(-\rho Y)} \leq \E{\exp(-\rho X)}, \; \forall \; \rho \geq 0 \;.
\end{align}
One useful property of LT ordered random variables is that for all {\cm} functions $g$, we have
\begin{align}
\label{J_CapOrd_eqn:LT_cm_order}
X \orderl{Lt} Y \iff \E{g(Y)} \leq \E{g(X)} .
\end{align}
In other words, the generator $\F$ can be enlarged to the set of all {\cm} functions without changing the stochastic order \cite{book:muller02}. Further, whenever $g \in \BF$, \eqref{J_CapOrd_eqn:LT_cm_order} holds with a reversal in the inequality. In a wireless communications context, let $\rho \geq 0$ be the average SNR, and $\rho X$, $\rho Y$ represent the instantaneous SNRs of two fading distributions. If $g(x)$ corresponds to the instantaneous symbol error rate $P_{\rm e}(\rho x)$ of a modulation scheme with {\cm} error rate function, then \eqref{J_CapOrd_eqn:LT_cm_order} can be used to obtain comparisons of averages of symbol error rates over pairs of fading channels, even in cases where a closed-form expression for the same is intractable. 


\subsection{Shannon Transform}
In what follows, we formally describe the Shannon transform, which is the basis of the proposed stochastic order in this paper. The Shannon transform of a nonnegative random variable $X$ is defined as \cite[pp. 44]{book:verdu}:
\begin{align}
\label{J_CapOrd_eqn:shannon_tf_def}
\Cerg{X}{} := \E{\C{\rho X}} \;, \rho  \geq 0 . 
\end{align}
Two new representations of $\Cerg{X}{}$, which are useful in this paper are now obtained. Using \eqref{J_CapOrd_eqn:cap_alt}, it is easy to show that \eqref{J_CapOrd_eqn:shannon_tf_def} can be represented as a Laplace transform, given by
\begin{align}
\label{J_CapOrd_eqn:shannon_tf_CM}
\Cerg{X}{} = \int\limits_{0}^{\infty} \exp(-u/\rho) \frac{1- \phi_{X}(u)}{u} \D u \;,
\end{align}
for $\rho > 0$, where $\phi_{X}(u) := \E{\exp(-u X)}, u >0$. Using \eqref{J_CapOrd_eqn:bernstein} with \eqref{J_CapOrd_eqn:shannon_tf_CM}, it is immediate that $\Cerg{X}{}$ is a {\cm} function of $1/\rho$. A second representation of $\Cerg{X}{}$ which can be derived from \eqref{J_CapOrd_eqn:shannon_tf_CM} shows that $\Cerg{X}{}$ is also the Stieltjes transform \cite[p. 325]{book:widder46} of the complimentary CDF of $X$, when evaluated at $1/\rho$:
\begin{align}
\label{J_CapOrd_eqn:shannon_tf_ST}
\Cerg{X}{} = \int\limits_{0}^{\infty} \frac{1- \CDF{X}{u}}{(1/\rho + u)} \D u \;,
\end{align} 
where $\rho >0$. Representation \eqref{J_CapOrd_eqn:shannon_tf_ST} is used in proving some properties of the ergodic capacity order discussed in Section \ref{J_CapOrd_sec:props}. Additionally, \eqref{J_CapOrd_eqn:shannon_tf_ST} permits us to comment on the convergence of $\Cerg{X}{}$:
\begin{proposition}
\label{J_CapOrd_prop:conv1}
If $\Cerg{X}{}$ exists for any $\rho \in (0, \infty)$, then $\Cerg{X}{}$ exists for every $\rho \in (0, \infty)$.
\end{proposition}
\begin{IEEEproof}
From \eqref{J_CapOrd_eqn:shannon_tf_ST}, it is seen that $\Cerg{X}{}$ is the Stieltjes transform of a real valued function. If the Stieltjes transform of a function exists at any point on $\re^{+}$, then it exists at all points on $\re^{+}$ \cite[p. 326]{book:widder46}. This completes the proof.
\end{IEEEproof}
We now provide examples of random variables for which the ergodic capacity is finite for $\rho < \infty$ using the following proposition:
\begin{proposition}
\label{J_CapOrd_prop:ex1}
Let $\CDF{X}{\cdot}$ denote the cumulative distribution function of a RV $X$. If for some $\delta \in (0,1]$, $\int_{0}^{t} 1 - \CDF{X}{u} \D u =  O(t^{1-\delta}), t \rightarrow \infty$, then $\Cerg{X}{} < \infty$.
\end{proposition}
\begin{IEEEproof}
First, observe that $\int_{0}^{\infty}(s+t)^{-1}\D \alpha(t)$ exists if $\alpha(t) = O(t^{1-\delta}), t \rightarrow \infty$, for some $\delta >0$ \cite[p. 330 (Theorem 3b)]{book:widder46}. The proposition then follows by letting $\alpha(t) = \int_{0}^{t} 1 - \CDF{X}{u} \D u$. This completes the proof.
\end{IEEEproof}
In Proposition \ref{J_CapOrd_prop:ex1}, the case of $\delta = 1$ is equivalent to the condition that the mean of $X$ is finite. It is therefore straightforward to see that the ergodic capacity of fading distributions such as Nakagami-$m$ and Rician is finite at all finite SNR, since these distributions have finite average power. We now proceed to define a stochastic order for comparing fading distributions based on the Shannon transform. 
\section{The Ergodic Capacity Order}
\label{J_CapOrd_sec:erg_cap_ord}
Recall that the ergodic capacity of a single-input single-output (SISO) system is given by $\E{\C{\rho X}}$, where $X$ is the square of the amplitude of the complex fading gain, and is defined as the instantaneous fading power of the channel. It is straightforward to see through an application of Jensen's inequality that the AWGN channel (with no fading) outperforms every fading distribution with same average channel power, in terms of the ergodic capacity at all SNR. However, given two fading distributions, it is not trivial to compare them based on the ergodic capacity, as obtaining a closed-form expression for the ergodic capacity of many fading channels is analytically intractable. Motivated by this, we propose a stochastic ordering method, which can be used to compare the ergodic capacity of two different fading channels. Note that, in this paper, we represent the squared magnitudes of the fading coefficients using the alphabets $X$, $Y$. This differs from the the convention of some authors, who denote the input symbol using $X$ and the output symbol using $Y$.
\subsection{Definition}
\begin{defn}
\label{J_CapOrd_eqn:cap_ord_def}
If $X$ and $Y$ are arbitrary nonnegative RVs, then $X$ is said to be dominated by $Y$ in the ergodic capacity order (i.e. $X \orderl{c} Y$), if the Shannon transforms of $X$ and $Y$ exist and $\Cerg{X}{} \leq \Cerg{Y}{}$ for $\rho \geq 0$.
\end{defn}
For this stochastic order, the generator is chosen as $\F = \lbrace g(x) : g(x) = \C{\rho x},  \rho \geq 0 \rbrace$. Distributions of interest for which the ergodic capacity is finite at all finite SNR can be determined using either Proposition \ref{J_CapOrd_prop:conv1} or Proposition \ref{J_CapOrd_prop:ex1}. Next, some useful properties of the capacity order and a few examples of ergodic capacity ordered RVs are discussed.
\subsection{Properties}
\label{J_CapOrd_sec:props}
The following properties hold for nonnegative RVs.
\begin{enumerate}
\item[S1:] $X \orderl{c} Y \iff \E{g(X)} \leq \E{g(Y)}$, $\forall g \in $ {\TBF}, such that the expectations exist.
\item[S2:] $X \orderl{c} Y \iff g(X) \orderl{c} g(Y)$, $\forall g \in \cTBF$.
\item[S3:] $X \orderl{Lt} Y \implies X \orderl{c} Y$.
\item[S4:] Let $X_{1},\ldots,X_{M}$ independent and $Y_{1},\ldots,Y_{M}$ independent. If $X_{m} \orderl{c} Y_{m} , \;  m = 1,\ldots,M$, then $g\left(X_{1},\ldots,X_{M}\right) \orderl{c} g\left(Y_{1},\ldots,Y_{M}\right)$, $\forall g \in \mathcal{CTBF}_{M}$.
\item[S5:] If $X \orderl{c} Y$ and $Y \orderl{c} Z$, then $X \orderl{c} Z$.
\item[S6:] If $X \orderl{c} Y$ and $Y \orderl{c} X$, then $\CDF{X}{\cdot} = \CDF{Y}{\cdot}$ a.e..
\end{enumerate}

The proofs of these properties follow as special cases of those presented in Appendix \ref{J_CapOrd_appendix:MIMOproperties}. A straightforward implication of Property S1 is that if $X \orderl{c} Y$, then $\E{X} \leq \E{Y}$, since $g(x) =x$ is a Thorin-Bernstein function. In other words, if one fading channel has a higher ergodic capacity than another at all SNR, then it is necessary that the average fading power of the first channel is no smaller than that of the second. 
Properties S5 and S6 together constitute the definition of a partial order, and consequently $\orderl{c}$ is a partial order on nonnegative RVs.

Interpreting $\rho X$ and $\rho Y$ as the instantaneous SNRs of two different fading channels, Properties S1-S6 are useful in obtaining the conditions under which the ergodic capacity of a composite system with coding/decoding capabilities only at the transmitter/receiver under the channel $Y$ is greater than that under $X$ at all SNR. Although Property S3 suggests that every pair of Laplace transform ordered random variables also obey the ergodic capacity order, the converse is not true in general. A counterexample can be found in \cite{paper:adithya11, paper:AdithyaCapacity12}. Thus, it is possible that the average symbol error rate of differential binary phase shift keying modulation in channel $X$ is less than that in $Y$ at high SNR, while the situation reverses when the capacity achieving code is applied on both channels. Interpreting the ergodic
capacity as what is achievable by coding over an i.i.d. time-extension of the channel, we reach
the conclusion that even though $Y$ offers more diversity than $X$ for an uncoded system, the i.i.d.
extension of $X$ lends itself to more diversity than that of $Y$. To put it more simply, at high SNR,
it is possible for one fading channel to be superior to another in terms of error rates in the absence of
coding, while being inferior when the capacity achieving code is employed over both channels. 

\subsection{Examples}
Next, we give examples of pairs of RVs $X,Y$ relevant to wireless communications, for which $X \orderl{c} Y$ holds. In general, establishing ergodic capacity ordering using its definition is often inconclusive, since the corresponding integrals are intractable. Fortunately, using Property S3, it is possible to provide examples of pairs of RVs which obey capacity ordering. In what follows, examples of parametric fading distributions which obey the ergodic capacity order are given. These distributions are also known to satisfy the Laplace transform order \cite{paper:adithya11}.
\subsubsection{Nakagami Fading}
The Nakagami-$m$ fading model, for which the envelope $\sqrt{X}$ is Nakagami distributed, and the instantaneous fading power $X$ is Gamma distributed, with PDF given by 
\begin{align}
\label{J_CapOrd_eqn:gamma_PDF} 
\pdf{X}{x} = \frac{m^{m}}{\Gamma(m)}x^{m-1}\exp(-mx) \; , x \geq 0 \;,
\end{align}
where $m>0$ is the line of sight parameter, and $\Gamma(r) :=  \int_{0}^{\infty} t^{r-1} \exp(-t) \D{t}$ is the gamma function. Let $X \sim {\rm Gamma}(m^{X})$, and $Y \sim {\rm Gamma}(m^{Y})$ with $m^{X} \leq m^{Y}$. For this case, it is easy to verify that $X \orderl{Lt} Y$, which implies that $X \orderl{c} Y$, according to Property S3. Property S3 requires the existence of the Shannon transforms, which is proved as follows. Observing that $\E{X}=\E{Y}=1$ is finite, from Proposition \ref{J_CapOrd_prop:ex1}, the Shannon transforms exist. This is because setting $\delta = 1$ in Proposition \ref{J_CapOrd_prop:ex1} is equivalent to saying that the mean value is finite. 
\subsubsection{Rician Fading}
The Rician fading model: In this case, the envelope of the fading i.e., $\sqrt{X}$ is Rice distributed with line of sight parameter $K$, and the corresponding instantaneous fading power distribution is given by  
\begin{equation}
\label{J_CapOrd_eqn:Ricean_PDF} 
 \pdf{X}{x} =  (K+1) \exp\left[-(K+1) x -K\right] I_{0}\left(2 \sqrt{K (K+1) x} \right) \; ,
\end{equation}
where $I_{0}(t) := \sum_{m=0}^{\infty} (t/2)^{2m}/(m! \Gamma(m+1)) $ is the modified Bessel function of the first kind of order zero. If the distribution of $X$ and $Y$ have parameters $K^{X}$ and $K^{Y}$ respectively, with $K^{X} \leq K^{Y}$, then $X \orderl{c} Y$. The existence of the Shannon transforms is established in way similar to that of the Nakagami-$m$ case. 
\subsubsection{Hoyt Fading}
The Nakagami-$q$ (Hoyt) fading model: Here, the envelope of the fading RV, given by $\sqrt{X}$ is Hoyt distributed, and the density of the (unit mean) instantaneous fading power is given by
\begin{align}
\label{J_CapOrd_eqn:hoyt_pdf}
\pdf{X}{x} = a \exp(-a^{2} x) I_{0}(b x)\;,
\end{align}
where $a = (1+q^{2})/2q$, $b = (1-q^{4})/4q^{2}$. If $X$ and $Y$ have parameters $q^{X}$ and $q^{Y}$ respectively, where $q^{X} \leq q^{Y}$, then $X \orderl{c} Y$. The existence of the Shannon transforms is established in way similar to that of the Nakagami-$m$ case. 

For the cases of Nakagami, Rician and Hoyt fading, the increase in ergodic capacity with increase in the LoS parameter of the distribution is not due to an increase in the average fading power, since $\E{X} = \E{Y}$, which is independent of the LoS parameter. 

In what follows, we show that ergodic capacity ordering of a given SISO system under two different fading channels can be used to make meaningful conclusions when a number of such systems are combined to form a system involving multiple random variables. 
\section{Systems Involving Multiple Random Variables}
\label{J_CapOrd_sec:sys_mult_rv}
In order to illustrate the applicability of the ergodic capacity order to compare the performance of systems, we provide examples of composite systems where ergodic capacity ordering of component SISO systems can be used to conclude the capacity ordering of the system, and also some applications where this is not necessarily the case. Such generic conclusions can be made even when closed form expressions for the ergodic capacity are not available. Throughout, we assume that the receiver has a perfect estimate of the instantaneous fading power, while the transmitter does not possess any such information.

\subsection{Diversity Combining Systems}

As examples of systems involving multiple fading links, we first consider diversity combining schemes such as maximum ratio combining (MRC) and equal gain combining (EGC) using $M$ receive antennas, for which we aim to compare the ergodic capacity under two different fading scenarios. Using the properties of the ergodic capacity order, we now show that diversity combining systems formed using a better set of components yields a system with a higher ergodic capacity, for the two schemes considered. 
\subsubsection{Maximum Ratio Combining}
Conditioned on the instantaneous fading power $X_{m} = x_{m}$, $m=1,\ldots,M$, the fading power after combining is given by 
\begin{align}
\label{J_CapOrd_eqn:snr_mrc}
g_{_{\rm MRC}}(x_{1},\ldots,x_{M}) = \sum_{m=1}^{M} x_{m} \;.
\end{align}
The ergodic capacity corresponding to this combining scheme is given by 
\begin{align}
\label{J_CapOrd_eqn:cerg_mrc}
\Cerg{X}{MRC} = \E{\C{\rho g_{_{\rm MRC}}(X_{1},\ldots,X_{M})}}\;.
\end{align}
It is easy to see that $\Cerg{Y}{MRC}$ is finite if the Shannon transforms of $Y_{m}$, $m=1,\ldots,M$ exist. We then obtain the following result, which can be used to compare the ergodic capacity of MRC in two different fading environments characterized by instantaneous fading powers $(X_{1},\ldots,X_{M})$ and $(Y_{1},\ldots,Y_{M})$:
\begin{proposition}
If $X_{m} \orderl{c} Y_{m}$, $m=1,\ldots,M$, then $\Cerg{X}{MRC} \leq \Cerg{Y}{MRC}$, at all $\rho \geq 0$.
\end{proposition}
\begin{IEEEproof}
We first verify that $g_{_{\rm MRC}}(\cdot)$ is a composable Thorin-Bernstein function. Then, we use Property S4 to conclude $\Cerg{X}{MRC} \leq \Cerg{Y}{MRC}$, at all $\rho \geq 0$, when $X_{m} \orderl{c} Y_{m}$, $m=1,\ldots,M$. 

To show that $g_{_{\rm MRC}}(\cdot) \in \cTBF$, treat $x_{1}$ in $g_{_{\rm MRC}}(\cdot)$ as the variable, while treating other arguments as constants, to get $g_{_{\rm MRC}}(x_{1};x_{2},\ldots,x_{M}) = x_{1}+k$, where $k=\sum_{m=2}^{M} x_{m}$. By definition, $g_{_{\rm MRC}} \in$ {\cTBF} if and only if $h_{\mathrm{MRC}}(x):=g_{_{\rm MRC}}'(x;x_{2},\ldots,x_{M})/g_{_{\rm MRC}}(x;x_{2},\ldots,x_{M}) = (x+k)^{-1}$ is a Stieltjes function. This is indeed the case, since $h_{\mathrm{MRC}}(\cdot)$ satisfies \eqref{J_CapOrd_eqn:st_nec_suf} with $a=0,b=0$, and $\mu(s) = \delta(s)$. Now, assuming $X_{m} \orderl{c} Y_{m}, m = 1,\ldots,M$, we have from Property S4 $g_{_{\rm MRC}}(X_{1},\ldots,X_{M}) \orderl{c} g_{_{\rm MRC}}(Y_{1},\ldots,Y_{M})$, which implies $\Cerg{X}{MRC} \leq \Cerg{Y}{MRC}$, at all $\rho \geq 0$. 
\end{IEEEproof}
Thus, if $Y_{m}$ dominates $X_{m}$ in the ergodic capacity order for $m=1,\ldots,M$, then the MRC system with fading links given by $Y_{1},\ldots,Y_{M}$ will have a higher ergodic capacity than that with $X_{1},\ldots,X_{M}$ at all SNR.
\subsubsection{Equal Gain Combining}
For the case of equal gain combining, the ergodic capacity is given by 
\begin{align}
\label{J_CapOrd_eqn:cerg_egc}
\Cerg{X}{EGC} = \E{\C{\rho g_{_{\rm EGC}}(X_{1},\ldots,X_{M})}}\;,
\end{align}
where $g_{_{\rm EGC}}(\cdot)$ represents the combined instantaneous fading power, and is given by 
\begin{align}
g_{_{\rm EGC}}(x_{1},\ldots,x_{M})= M^{-1}\left( \sum_{m=1}^{M} \sqrt{x_{m}}\right)^{2}\;.
\end{align}
It is possible to show that $\Cerg{Y}{EGC}$ is finite if the Shannon transforms of $Y_{m}$, $m=1,\ldots,M$ exist, by using the Cauchy-Schwarz inequality in addition to showing that the Shannon transform of $\sqrt{Y}_{m}$ exists if the Shannon transform of $Y_{m}$ exists. While closed-form expressions for the ergodic capacity of equal gain combining for several fading distributions are unknown, it is still possible for us to compare these quantities using the ergodic capacity ordering of component branches:
\begin{proposition}
\label{prop:EGC}
Let $X_{m} \orderl{c} Y_{m}$, $m=1,\ldots,M$. Then $\Cerg{X}{EGC} \leq \Cerg{Y}{EGC}$, at all $\rho \geq 0$.
\end{proposition}
\begin{IEEEproof}
We first prove that $g_{_{\rm EGC}} \in$ {\cTBF}, and then use Property S4 to complete the proof. In order to show that $g_{_{\rm EGC}} \in$ {\cTBF}, treat $x_{1}$ as the variable and all the other arguments of $g_{_{\rm EGC}}$ as constants, so that $g_{_{\rm EGC}}(x_{1};x_{2},\ldots,x_{M}) = M^{-1}(x_{1} + 2\sqrt{x_{1}}k + k^{2})$, where $k = \sum_{m=2}^{M}x_{m}$. 
By definition, $g_{_{\rm EGC}}(\cdot)$ in {\cTBF} if and only $h(x) := g_{_{\rm EGC}}'(x;k)/g_{_{\rm EGC}}(x;k) =(x+k\sqrt{x})^{-1}$ is a Stieltjes function. To show that $h \in \set{S}$, observe that $(h(x^{-1}))^{-1} = x^{-1} + kx^{-1/2}$ is a Stieltjes function, since any function of the form $x^{\alpha-1}, 0 \leq \alpha \leq 1$ is a Stieltjes function \cite[p. 13]{book:schilling}, and positive linear combinations of Stieltjes functions also yields a Stieltjes function. To complete the argument, since $(h(x^{-1}))^{-1} \in \set{S}$, $h(x)$ must also belong to $\set{S}$ \cite[p. 66]{book:schilling}. Consequently, $g_{_{\rm EGC}}(\cdot) \in \cTBF$. The rest of the proof follows arguments similar to the MRC case.
\end{IEEEproof}
Using Proposition \ref{prop:EGC}, we infer that if a collection of SISO systems with higher ergodic capacity is combined to form an EGC system, then the composite EGC system will have higher overall ergodic capacity.
\subsection{Multi-Hop Amplify-Forward Relay System}
\begin{figure}[htb]
\centering
\includegraphics[scale=0.5]{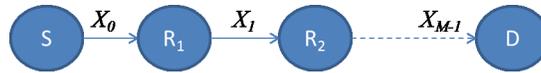}
\caption{$M$-hop relay. S represents the source, $\text{R}_{m}$ represent the relays and D represents the destination.}\label{J_CapOrd_fig:multihop_fig}
\end{figure}

We now turn our attention to multi-hop amplify-forward (MH-AF) relay systems. This is an example of a system where despite component-wise ergodic capacity ordering of individual hops, the overall system need not have a higher ergodic capacity at all SNR. The system consists of a source, which transmits data to a destination using $M-1$ half-duplex variable gain relays, which possess receive CSI (Figure \ref{J_CapOrd_fig:multihop_fig}). The source transmits in time slot $1$ to relay $1$, and relay $m$ in turn amplifies and retransmits to relay $m+1$ in time slot $m+1$, $m = 1,\ldots,M-2$, while relay $M-1$ amplifies and transmits to the destination in time slot $M$. The gain of the $m^{th}$ relay node is given by $\alpha_{m} = \rho / (\rho X_{m-1} +1)$ \cite{paper:alouini03}, where $X_{m}$ is the instantaneous fading power of the $m^{th}$ hop, for $m=1,\ldots,M-1$. $X_{0}$ denotes the instantaneous fading power of the channel between the source and the first relay node. It is assumed that coding/decoding capabilities are provided to the transmitter/receiver alone. In this case, the end-to-end ergodic capacity is given by
\begin{align}
\label{J_CapOrd_eqn:cap_MH-AF}
\Cerg{X}{MH-AF} = \E{\C{ g_{_{\rm MH-AF}}(X_{0},\ldots,X_{M-1})}}\;,
\end{align}
where $g_{_{\rm MH-AF}}(x_{0},\ldots,x_{M-1}):= (\prod_{m=0}^{M-1} [(1+(\rho x_{m})^{-1})]-1)^{-1}$.
Exact expressions for the ergodic capacity in arbitrary fading channels are intractable, even for the two-hop case. Previously, the ergodic capacity of such a relay in fading channels has been obtained as an infinite series in \cite{paper:waqar10}. Nevertheless, even in the absence of closed-form expressions, it is possible to compare the ergodic capacities of two such relay networks which are identical, except for the fading distribution across the hops. 

In order to compare the performance of the MH-AF relay in two different fading scenarios, let $X_{m}$ and $Y_{m}$ denote the instantaneous fading power of the $m^{th}$ link of the first and second fading channels respectively, for $m=0,\ldots,M-1$. 
\begin{proposition}
If $X_{m} \orderl{Lt} Y_{m}$, $m=0,\ldots,M-1$, then $\Cerg{X}{MH-AF} \leq \Cerg{Y}{MH-AF}$ at all $\rho \geq 0$.
\end{proposition}
\begin{IEEEproof}
To establish this result, we recall that a property similar to Property S4 holds for LT ordered random variables: If $X_{m} \orderl{Lt} Y_{m}, m = 0,\ldots M-1$, then $g(X_{0},\ldots,X_{M-1}) \orderl{Lt} g(Y_{0},\ldots,Y_{M-1})$, whenever $g$ which is a Bernstein function in each variable, while viewing all the other variables as constants \cite[Theorem 5.A.7]{book:shaked}. Now, this can be established by straight-forward differentiation with respect to $x_{i}$. As a result, if the instantaneous fading powers satisfy $X_{m} \orderl{Lt} Y_{m}, m=0,\ldots,M-1$, then $g_{_{\rm MH-AF}}(X_{0},\ldots,X_{M-1}) \orderl{Lt} g_{_{\rm MH-AF}}(Y_{0},\ldots,Y_{M-1})$, and therefore from Property S3, we have $g_{_{\rm MH-AF}}(X_{0},\ldots,X_{M-1}) \orderl{c} g_{_{\rm MH-AF}}(Y_{0},\ldots,Y_{M-1})$. The proposition then follows, since ergodic capacity ordered RVs have ordered expectations.
\end{IEEEproof}
In other words, if each hop of $Y$ dominates the corresponding hop of $X$ in the Laplace transform order, then the overall ergodic capacity of the $M$-hop MH-AF relay formed using $\lbrace Y_{m} \rbrace_{m=0}^{M-1}$ will be higher than that formed using $\lbrace X_{m} \rbrace_{m=0}^{M-1}$.

\begin{figure}[htb]
\centering
\includegraphics[height=7.0cm,width=9cm]{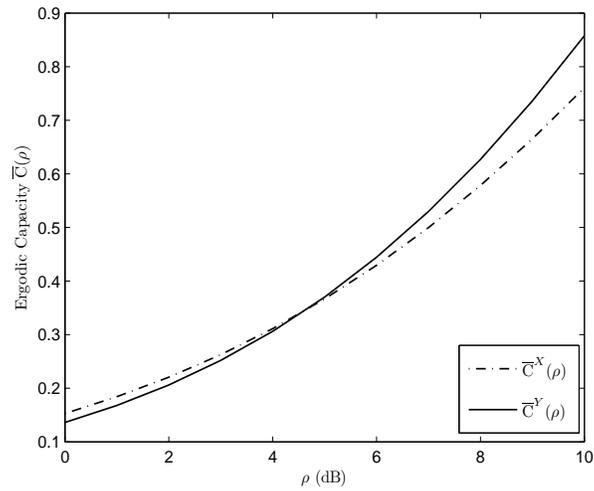}
\caption{Ergodic capacity of amplify-forward relay with $M=3$ slots. The instantaneous SINR is Pareto distributed with parameters $\beta_{X} = 1$ (dashed line) and $\beta_{Y} = 3$ (solid line).}\label{J_CapOrd_fig:mh_af_pareto}
\end{figure}

However, this conclusion does not hold if we make the weaker assumption that $X_{m} \orderl{c} Y_{m}$, instead of $X_{m} \orderl{Lt} Y_{m}$, $m=0,\ldots,M-1$. In other words, componentwise ordering of links in the ergodic capacity ordering sense does not imply the ordering of the overall system. To see a counterexample, consider the case of an interference dominated channel, where the instantaneous fading power to interference power ratio $X_{m}$ are independent and Pareto-type distributed with parameter $\beta_{X}$ \cite{pun07}:  
\begin{align}
\CDF{X_{m}}{x} = \frac{x^{\beta_{X}}}{1+x^{\beta_{X}}}, x>0, \beta_{X} >0 \;,
\end{align}
and $Y_{m}$ similarly with parameter $\beta_{Y}$, where $\beta_{X} \leq \beta_{Y}$. In this case, it can be shown that $X_{m} \orderl{c} Y_{m}$, but $X_{m} \norderl{Lt} Y_{m}$, $m=0,\ldots,M-1$. As an illustrative example, Fig. \ref{J_CapOrd_fig:mh_af_pareto} shows the numerically evaluated ergodic capacities of a multi-hop relay with $M=3$ hops under Pareto-type distributed signal-to-interference ratio with parameters $\beta_{X}=1$ and $\beta_{Y}=3$, so that for each hop $X_{m} \orderl{c} Y_{m}, m=0,1,2$ is satisfied. It is observed from Fig. \ref{J_CapOrd_fig:mh_af_pareto} that for $\rho < \rho_{0}$, where $\rho_{0} \approx 5$ dB, $X$ is a better channel than $Y$ in the ergodic capacity order, while for $\rho \geq \rho_{0}$, the situation is reversed. 

In summary, the MH-AF system is an example of a case where contrary to intuition, it is possible for a fading channel system $\lbrace Y_{m} \rbrace_{m=0}^{M-1}$ to not have a higher ergodic capacity at all SNR than that of $\lbrace X_{m} \rbrace_{m=0}^{M-1}$, even though the ergodic capacity of each $Y_{m}$ is higher than that of $X_{m}$, $m=0,\ldots,M-1$ at all SNR. 
\subsection{Fading Multiple Access Channel}
\label{J_CapOrd_sec:mac}
In this example, we focus on comparing the ergodic capacity regions of a multi-user Gaussian MAC network in two different fading scenarios. Consider the following system model:
\begin{align}
r = \sqrt{\rho} \sum\limits_{m=1}^{M}h_{m} s_{m} + v \;,
\end{align}
where $r$ is the received signal, $\rho$ is the average SNR of each user, $s_{m}$ is the transmitted symbol of user $m$, $h_{m}$ is the complex i.i.d (across time) ergodic fading between each user and the destination, and $v$ is the AWGN at the receiver. It is assumed that only the receiver possesses CSI of all the users. The receiver intends to decode the signals from all the users. If $X_{m} := |h_{m}|^{2}, m=1,\ldots,M$, then the ergodic capacity region $\Cerg{\cdot}{MAC}$ is the set of all rate $M$-tuples that satisfy \cite[pp. 407]{book:cover},
\begin{align}
\label{J_CapOrd_eqn:mac_cap_region}
\sum\limits_{m \in \mathcal{S}}R_{m}(\rho ) \leq \E{\C{\rho \sum_{\mathcal{S}}X_{m} }} \;,
\end{align}
where $\mathcal{S} \subset 2^{\lbrace 1,\ldots, M \rbrace }$. Using the ergodic capacity order, we can now make the following observation which links the ordering of ergodic capacities of each user to the overall ergodic capacity region of the fading MAC.
\begin{proposition}
If $X_{m} \orderl{c} Y_{m}, m = 1,\ldots,M$, then $\Cerg{X}{MAC} \subseteq \Cerg{Y}{MAC}$, for $\rho \geq 0$.
\end{proposition}
\begin{IEEEproof}
To begin with, observe that $g_{_{{\rm MAC}, \mathcal{S}}}(x_{1},\ldots,x_{M}):= \sum_{\mathcal{S}}x_{m}$ belongs to $\cTBF_{\card{\set{S}}}$. Now, if $X_{m} \orderl{c} Y_{m}, m = 1,\ldots,M$, from Property S4 it follows that 
\begin{align}
g_{_{{\rm MAC}, \mathcal{S}}}(X_{1},\ldots,X_{M}) \orderl{c} g_{_{{\rm MAC}, \mathcal{S}}}(Y_{1},\ldots,Y_{M}),\forall \mathcal{S} \subset 2^{\lbrace 1,\ldots, M \rbrace }\;.
\end{align}
Hence, if $X_{m} \orderl{c} Y_{m}, m = 1,\ldots,M$, then $\Cerg{X}{MAC} \subseteq \Cerg{Y}{MAC}$, for all $\rho \geq 0$.
\end{IEEEproof}
In other words, if each user of the system $X$ has a higher ergodic capacity than the corresponding user in the system $Y$, then $\Cerg{X}{MAC} \subseteq \Cerg{Y}{MAC}$, for $\rho  \geq 0$. 

\section{MIMO Ergodic Capacity Order}
\label{J_CapOrd_sec:MIMO}
In this section, the ergodic capacity ordering of MIMO systems is presented. Some properties of this stochastic order are discussed, and an application of this framework in a MIMO MAC setting is presented. Before doing so, we formally define a MIMO system through its single letter characterization:
\begin{align}
\label{J_CapOrd_eqn:mimo_sys_model}
\vect{r} = \sqrt{\rho}\mat{H} \vect{s} + \vect{v} \;,
\end{align}
where $\vect{r}$ is the received signal, $\mat{H}$ is a complex $N_{R} \times N_{T}$ random matrix which captures the effect of ergodic quasi-static fading, $\vect{v} \sim \mathcal{CN}(\vect{0},\mat{I})$ is the additive noise, $\vect{s}$ is the transmitted symbol vector, and $\rho$ is the average SNR per transmit antenna. $\mat{H}$ and $\vect{v}$ are assumed to be i.i.d across time, as a result of which a time index has not been used in \eqref{J_CapOrd_eqn:mimo_sys_model}. Further, it is assumed that the receiver tracks the channel fading realizations $\mat{H}$, while no such CSI is available at the transmitter. For this system model, the instantaneous fading power is given by $\mat{H}^{\rm H}\mat{H}$, and is denoted as $\mat{X}$. In this case, the ergodic capacity is the Shannon transform of the instantaneous fading power, and is given by $\Cerg{X}{MIMO} = \E{\LD{\rho \mat{X}}}$. 
\\
Remark: The Shannon transform for an arbitrary distribution on positive semidefinite matrices need not exist. Using Proposition \ref{J_CapOrd_prop:ex1}, it can be shown that the Shannon transform for a positive semidefinite matrix $\mat{X}$ exists, if there exists some $\delta \in (0,1]$, such that $\int_{0}^{t} \E{ 1- \CDF{\eig{Q}{X}}{u}} \D u = O(t^{1-\delta}), t \rightarrow \infty$, where $Q$ is uniformly picked from $\lbrace 1,\ldots, n \rbrace$.

In what follows, we define a partial order on the instantaneous fading power, which can be used to compare the ergodic capacity of composite MIMO systems under two different fading environments.
\subsection{Definition and Properties}
\begin{defn}
\label{def:MIMO_cap_order}
For two random positive semidefinite matrices $\mat{X}$, $\mat{Y}$, we say that $\mat{X}$ is dominated by $\mat{Y}$ in the MIMO ergodic capacity order, and write $\mat{X} \orderm{c} \mat{Y}$, if the Shannon transforms of $\mat{X}$ and $\mat{Y}$ exist and $\E{\Tr{ \mlog \left( \mat{I} + \rho \mat{X}\right)}} \leq \E{\Tr{ \mlog \left( \mat{I} + \rho \mat{Y}\right)}}$, for all $  \rho \geq 0$.
\end{defn} 
In Definition \ref{def:MIMO_cap_order}, $\mathbf{\mlog}(\cdot)$ is to be viewed as a matrix function, in the sense of Section \ref{J_CapOrd_subsec:mat_fun}. It is easy to show that $\mat{X} \orderm{c} \mat{Y}$ is equivalent to $\E{\LD{\rho \mat{X}}} \leq \E{\LD{\rho \mat{Y}}}$, at all $\rho \geq 0$. In contrast to the ergodic capacity order on random variables, the MIMO ergodic capacity corresponding to two different random matrices $\mat{X}$ and $\mat{Y}$ may be identical (for example, when $\mat{Y} = \mat{UXU}^{\rm H}$, where $\mat{U}$ is a unitary matrix). In this circumstance, we write $\mat{X} =_{c} \mat{Y}$. In what follows, some properties of the MIMO ergodic capacity order are developed, which can be viewed as matrix analogues to the properties developed in Section \ref{J_CapOrd_sec:props}. The following properties are true for positive semi-definite random matrices, for which the Shannon transforms exist.
\begin{enumerate}
\item[M1:] If $\mat{X}, \mat{Y} \in \PSD{n}$, then $\mat{X} \orderm{c} \mat{Y} \iff \E{\Tr{ g(\mat{X})}} \leq \E{\Tr{ g(\mat{Y})}}$, for all $g: \re \rightarrow \re $, such that $ g \in {\TBF}$, provided the expectations exist.
\item[M2:] If $\mat{X}, \mat{Y} \in \PSD{n}$, then $\mat{X} \orderm{c} \mat{Y} \iff g(\mat{X}) \orderm{c} g(\mat{Y})$, for all $g: \re \rightarrow \re$, such that $ g \in {\cTBF}$.
\item[M3:] If $\mat{X}, \mat{Y} \in \PSD{n}$ and $\E{\Tr {\exp(-\rho \mat{X})}} \geq  \E{\Tr {\exp(-\rho \mat{Y})}} \; \forall \rho \geq 0$ then $\mat{X} \orderm{c} \mat{Y}$.
\item[M4:] Let $\lbrace \mat{X}_{m} \rbrace_{m=1}^{M}$, $\lbrace \mat{Y}_{m} \rbrace_{m=1}^{M}$ be independent random matrices in $\PSD{n}$, such that $\mat{X}_{m} \orderm{c} \mat{Y}_{m}$, $m=1,\ldots,M$. Let $g(\mat{X}_{1:M}) := g(\mat{X}_{1},\ldots,\mat{X}_{M})$, i.e., $g$ operates on $M$ $\PSD{n}$ matrices and produces a $\PSD{n}$ matrix. If $g: \re^{M} \rightarrow \re$ is such that $g \in \mathcal{CTBF}_{M}$ then $g(\mat{X}_{1:M}) \orderm{c} g(\mat{Y}_{1:M})$.
\item[M5:] If $\mat{X} \orderm{c} \mat{Y}$, and $\mat{Y} \orderm{c} \mat{Z}$, then $\mat{X} \orderm{c} \mat{Z}$.
\item[M6:] $\mat{X} =_{c} \mat{Y}$ if and only if $\sum_{i=1}^{n} \CDF{\eig{i}{\mat{X}}}{u} = \sum_{i=1}^{n} \CDF{\eig{i}{\mat{Y}}}{u}$, where $\CDF{\eig{i}{\mat{X}}}{\cdot}$ is the marginal CDF of the $i^{th}$ largest eigenvalue of $\mat{X}$.
\end{enumerate}

The proofs of properties M1-M4, and M6 can be found in Appendix \ref{J_CapOrd_appendix:MIMOproperties}, while Property M5 is straight-forward to establish, and its proof is omitted. 
Property M3 provides a useful sufficient condition to verify if two random matrices obey the MIMO ergodic capacity order. This is because $\E{\Tr {\exp(-\rho \mat{X})}} \geq  \E{\Tr {\exp(-\rho \mat{Y})}}$ at all $\rho \geq 0$ is equivalent to $\sum_{i=1}^{n} \E{\exp(-\rho \lambda_{i}(\mat{X}))} \geq \sum_{i=1}^{n} \E{\exp(-\rho \lambda_{i}(\mat{Y}))}, \forall \rho \geq 0$, and Laplace transforms of the eigenvalue distributions are more easy to compute, when compared to the expectations of the log-determinants. 

Next, we form an interesting interpretation of Property M6. From Property M6, it follows that $\mat{X} =_{c} \mat{Y}$ if and only if $\mathbb{E}_{Q}[\CDF{\eig{Q}{\mat{X}}}{u}] = \mathbb{E}_{Q}[\CDF{\eig{Q}{\mat{Y}}}{u}]$, where $Q$ is uniformly picked from $\lbrace 1,n \rbrace$. In other words, if the distribution of an eigenvalue picked randomly and uniformly from both matrices is identical, then the two random matrices are regarded to be the same with respect to the MIMO ergodic capacity order.

Although the proposed definition of the MIMO ergodic capacity order is one of many different possible partial orders on matrices, we assert that it is a natural generalization of the ergodic capacity order defined in Section \ref{J_CapOrd_sec:erg_cap_ord}. This is also elucidated by the fact that the properties M1-M3 and M5 are indeed straight-forward matrix generalizations of properties S1-S3 and S5 respectively. Further, the MIMO ergodic capacity order bears the following connection with the ergodic capacity order defined for random variables:
\begin{proposition}
\label{J_CapOrd_J_cap12:thm:SISO_MIMO}
Let $\eig{Q}{\mat{X}} \orderl{c} \eig{Q}{\mat{Y}}$, where $\eig{Q}{\mat{X}}$ is an eigenvalue of $\mat{X}$ picked uniformly from the set of eigenvalues of $\mat{X}$. Then $\mat{X}  \orderm{c}  \mat{Y}$. Conversely, if $\mat{X}  \orderm{c}  \mat{Y}$, then $\eig{Q}{\mat{X}} \orderl{c} \eig{Q}{\mat{Y}}$.
\end{proposition}

Given two MIMO fading systems $\mat{X}$ and $\mat{Y}$, Proposition \ref{J_CapOrd_J_cap12:thm:SISO_MIMO} implies that $\mat{Y}$ dominates $\mat{X}$ in the MIMO ergodic capacity order, if and only if a uniformly randomly selected eigen-channel of $\mat{Y}$ has a larger ergodic capacity than that of a uniformly randomly selected eigen-channel of $\mat{X}$.
\subsection{Application}
\label{J_CapOrd_J_cap12:sec:MIMOAppln}
An illustrative example to elucidate the efficacy of the MIMO ergodic capacity order is the $M$ user Gaussian MIMO-MAC, where user $i$ possesses $N_{t}$ antennas. We assume that only the receiver has CSI, and that each antenna of each user transmits independent signals. Further, each user is allocated the same transmit power $\rho$ per transmit antenna. In this case, the ergodic capacity region $\cmimomac{}(\rho)$ is given by \cite{paper:cioffi01}:
\begin{align}
\cmimomac{X}(\rho) :=& \left\lbrace (R_{1},\ldots , R_{M}) : \sum\limits_{i \in \set{S}}R_{i} \leq \E{\LD{ \rho g_{_{\rm MIMO-MAC, \set{S}}}(\mat{X}_{1:M})}} ,\right. \nonumber \\
&\left. \forall \set{S} \subseteq {\lbrace 1,\ldots ,M \rbrace} \right\rbrace \;, 
\end{align}
where $g_{_{\rm MIMO-MAC, \set{S}}}(\mat{X}_{1:M}) := \sum\limits_{i \in \set{S}}  \mat{X}_{i}$, with $\set{S} \subseteq {\lbrace 1,\ldots,M \rbrace}$. 
Clearly, when $\mat{X}_{i}$ is assumed to be the variable while viewing all other arguments of $g_{_{\rm MIMO-MAC, \set{S}}}(\cdot)$ as constant matrices, it can be seen that $g_{_{\rm MIMO-MAC, \set{S}}}(\cdot)$ is a Thorin-Bernstein matrix function of $\mat{X}_{i}$, for $i=1,\ldots,M$. Therefore, through property M4, $g_{_{\rm MIMO-MAC, \set{S}}}(\mat{X}_{1:M}) \orderm{c} g_{_{\rm MIMO-MAC, \set{S}}}(\mat{Y}_{1:M})$, whenever $\mat{X}_{i} \orderm{c} \mat{Y}_{i}, i = 1,\ldots,M$. Consequentially, $\cmimomac{X}(\rho) \subseteq \cmimomac{Y}(\rho)$, for $\rho \geq 0$. 
\section{Conclusion}
The ergodic capacity order and its properties can be exploited to obtain comparisons of ergodic capacities of composite systems across two different fading channels whose instantaneous SNRs satisfy the ergodic capacity order. For systems such as MRC and EGC which involve multiple instantaneous SNR RVs, we conclude that combining a better set of channels (in the ergodic capacity order) produces a system with a higher ergodic capacity. This conclusion is true for all systems whose end-to-end instantaneous SNR belongs to the $\cTBFm$ set. For systems whose end-to-end SNR does not belong to $\cTBFm$, component-wise ergodic capacity ordering of instantaneous SNR need not produce a system with a higher ergodic capacity. An example to illustrate this point is the MH-AF relay for which the instantaneous SINR is Pareto-type distributed. An extension of the ergodic capacity order to MIMO systems is also proposed herein. The properties of the ergodic capacity order can be used to compare the capacity regions of systems such as the multi-user MAC in two different fading environments, for both the single and multiple antenna case.

\appendices

\section{Proofs: Properties of MIMO Ergodic Capacity Order}
\label{J_CapOrd_appendix:MIMOproperties}
We now discuss the proofs of the properties of the MIMO ergodic capacity order. The proofs of the properties S1-S6 of the ergodic capacity order (for scalar RVs) are special cases of Properties M1-M6 respectively, and can be obtained by setting $n=1$. 
\subsection*{Proof of Property M1} 
Assume $\mat{X} \orderm{c} \mat{Y}$. Using the identity $\Det{\mat{I}+ \rho \mat{X}} = \prod_{i=1}^{n} \left( 1+ \rho \lambda_{i}(\mat{X}) \right)$, we can write
\begin{align}
\label{J_CapOrd_eqn:app2_M1_0}
\mat{X} \orderm{c} \mat{Y} \iff \E{\sum\limits_{i=1}^{n}\log (1+\rho \lambda_{i}(\mat{X}))} \leq \E{\sum\limits_{i=1}^{n}\log (1+\rho \lambda_{i}(\mat{Y}))} \;, \forall \rho >0.
\end{align}
Multiplying \eqref{J_CapOrd_eqn:app2_M1_0} by $\rho^{-1}$, and taking the limit as $\rho \rightarrow 0$, it is seen that 
\begin{align}
\label{J_CapOrd_eqn:app2_M1_1half}
\mat{X} \orderm{c} \mat{Y} \implies \E{\Tr{\mat{X}}} \leq \E{\Tr{\mat{Y}}}\;,
\end{align}
provided the Shannon transforms of $\mat{X}$ and $\mat{Y}$ exist, and $\E{\Tr{\mat{X}}}< \infty$ and $\E{\Tr{\mat{Y}}}< \infty$. 

It now follows from \eqref{J_CapOrd_eqn:app2_M1_0} and \eqref{J_CapOrd_eqn:app2_M1_1half} that
\begin{align}
\label{J_CapOrd_eqn:app2_M1_1}
& \mat{X} \orderm{c} \mat{Y}  \iff \nonumber \\
&\E{\sum\limits_{i=1}^{n} \log(1+t \rho \lambda_{i}(\mat{X}))\mu(t) + a + b \lambda_{i}(\mat{X})}  \leq \E{\sum\limits_{i=1}^{n} \log(1+t \rho \lambda_{i}(\mat{Y}))\mu(t) + a + b \lambda_{i}(\mat{Y})} ,\nonumber \\
& \forall a,b \geq 0, \mu(t) \geq 0, \rho >0, t >0 \;.
\end{align}
Integrating the right hand side of \eqref{J_CapOrd_eqn:app2_M1_1} over $t$ in the interval $[0,\infty)$ preserves the inequality in \eqref{J_CapOrd_eqn:app2_M1_1}. Therefore,
\begin{align}
\label{J_CapOrd_eqn:app2_M1_2}
\mat{X} \orderm{c} \mat{Y} & \implies \E{\sum\limits_{i=1}^{n} a+ b \lambda_{i}(\mat{X})+ \int\limits_{0}^{\infty} \log(1+t \rho \lambda_{i}(\mat{X}))\mu(t) \D t} \nonumber \\
& \leq \E{\sum\limits_{i=1}^{n} a+ b \lambda_{i}(\mat{Y})+ \int\limits_{0}^{\infty} \log(1+t \rho \lambda_{i}(\mat{Y}))\mu(t) \D t }, \forall \rho >0 \;.
\end{align}
The summand in \eqref{J_CapOrd_eqn:app2_M1_2} is an arbitrary Thorin-Bernstein function, since $a,b,\mu$ are arbitrary and nonnegative. Denoting this Thorin-Bernstein function by $g$, the direct part of the property is proved by observing from Section \ref{J_CapOrd_subsec:mat_fun} that $\E{\sum_{i=1}^{n}g(\eig{i}{\mat{X}})} = \E{\Tr{g(\mat{X})}}$. To prove the converse, choose $g(\mat{A}) = \log (\mat{I}+\rho \mat{A})$.
\subsection*{Proof of Property M2} 
Let $\mat{X}, \mat{Y} \in \PSD{n}$, and $\mat{X} \orderm{c} \mat{Y}$. Let $\phi: \re \rightarrow \re $ belong to {\TBF}, and $g: \re \rightarrow \re$ belong to $\cTBF$. Using the definition of matrix functions, it is easy to see that $f(\mat{X}) := \phi(g(\mat{X})) \in$ {\TBF}. From Property M1, it is seen that $\mat{X} \orderm{c} \mat{Y} \iff \E{\Tr{\phi(g(\mat{X}))}} \leq \E{\Tr{\phi( g(\mat{Y}))}}$. In other words, $g(\mat{X}) \orderm{c} g(\mat{Y})$, which proves the direct part of the property. To see the converse, choose $f$ as the identity map.
\subsection*{Proof of Property M3} 
Let $\mat{X}, \mat{Y} \in \PSD{n}$, and $\mat{X} \orderm{c} \mat{Y}$. Using Frullani's formula \eqref{J_CapOrd_eqn:cap_alt}, it is evident that an equivalent condition to $\mat{X} \orderm{c} \mat{Y}$ is given by
\begin{align}
\label{J_CapOrd_eqn:app2_M3_1}
\mat{X} \orderm{c} \mat{Y} &\iff \nonumber \\
&\E{\int\limits_{0}^{\infty} \frac{e^{-s}}{s}\sum\limits_{i=1}^{n} \left( 1 - \exp(-\rho s \eig{i}{X})\right) \D s} \leq \E{\int\limits_{0}^{\infty} \frac{e^{-s}}{s}\sum\limits_{i=1}^{n} \left( 1 - \exp(-\rho s \eig{i}{Y} )\right) \D s}.
\end{align}
Commuting the expectation and integral in \eqref{J_CapOrd_eqn:app2_M3_1}, we get
\begin{align}
\mat{X} \orderm{c} \mat{Y} &\iff  \nonumber \\
& \int\limits_{0}^{\infty} \frac{e^{-s}}{s}\E{\sum\limits_{i=1}^{n}  \exp(-\rho s \eig{i}{X})} \D s \geq \int\limits_{0}^{\infty} \frac{e^{-s}}{s}\E{\sum\limits_{i=1}^{n}  \exp(-\rho s \eig{i}{Y})} \D s.
\end{align}
Therefore, if $\E{\sum_{i=1}^{n}\exp(-\rho \eig{i}{X})} \geq \E{\sum_{i=1}^{n}\exp(-\rho \eig{i}{X})}, \rho >0$, then $\mat{X} \orderm{c} \mat{Y}$. The property then follows through the observation that $\E{\sum_{i=1}^{n}\exp(-\rho \eig{i}{X})} = \E{\Tr{\exp(-\rho \mat{X})}}$.
\subsection*{Proof of Property M4} This property is proved using mathematical induction. To begin with, choose a matrix function $\phi \in $ {\TBF}, and $\mat{X}_{1:m}:= [\mat{X}_{1},\ldots,\mat{X}_{m}]$ have independent and nonnegative random matrices as components. Assume likewise for $\mat{Y}_{1:m}:= [\mat{Y}_{1},\ldots,\mat{Y}_{m}]$. Now, for $m=1$, Property M4 is true due to Property M2. Next, let us assume Property M4 to be true for sequences of length $m-1$. Thus, for $g \in \cTBFm$ we have $g([\mat{C} \; \mat{X}_{1:m-1}]) \orderm{c} g([\mat{C} \; \mat{Y}_{1:m-1} ])$, where $g([\mat{C} \; \mat{X}_{1:m-1}]) := g(\mat{C},\mat{X}_{1},\ldots,\mat{X}_{m-1})$, and $\mat{C} \in \PSD{n}$. This implies
\begin{align}
\label{J_CapOrd_eqn:app2_M4_1}
\E{\Tr{\phi\left( g([\mat{C} \; \mat{X}_{1:m-1} ])\right)}} \leq \E{\Tr{\phi\left( g([\mat{C} \; \mat{Y}_{1:m-1} ])\right)}} \; ,
\end{align}
where we have used Lemma \ref{lem:multivMatrxFuntrace} and Lemma \ref{lem:multivMatrixFunComp}. Next, for sequences of length $m$, consider 
\begin{align}
\label{J_CapOrd_eqn:app2_M4_2}
\E{\Tr{\phi\left( g(\mat{X}_{1:m})\right)} | \mat{X}_{1} = \mat{C}}  &= \E{\Tr{\phi\left( g([\mat{C} \;\mat{X}_{2:m}])\right)}}\\
\label{J_CapOrd_eqn:app2_M4_3}
 \leq \E{\Tr{\phi\left( g([\mat{C}\;\mat{Y}_{2:m}])\right)}} & =  \E{\Tr{\phi\left( g(\mat{Y}_{1:m})\right)} | \mat{Y}_{1} = \mat{C}} \;,
\end{align}
where \eqref{J_CapOrd_eqn:app2_M4_3} follows from \eqref{J_CapOrd_eqn:app2_M4_2} due to \eqref{J_CapOrd_eqn:app2_M4_1}. Now, taking the expectation with respect to $\mat{X}_{1}$ on the left hand side of \eqref{J_CapOrd_eqn:app2_M4_2} and the right hand side of \eqref{J_CapOrd_eqn:app2_M4_3}, we get $\E{\Tr{\phi\left( g(\mat{X}_{1},\ldots,\mat{X}_{m})\right)}} \leq \E{\Tr{\phi\left( g(\mat{Y}_{1},\ldots, \mat{Y}_{m})\right)}}$. Since in the above argument, $\mat{X}_{1}$ is an indeterminate parameter, the same line of reasoning applies when conditioning on any other parameter, and the proof of the property thus follows.
\subsection*{Proof of Property M6}
To prove this property, let $\mat{X}, \mat{Y} \in \PSD{n}$, and $\E{\LD{\rho \mat{X} }} = \E{\LD{\rho \mat{Y} }}$. Using the representation of the log-determinant in terms of the eigenvalues, and \eqref{J_CapOrd_eqn:shannon_tf_ST}, it is seen that
\begin{align}
\label{J_CapOrd_eqn:app2_M6_1}
\mat{X} =_{c} \mat{Y} \iff \int\limits_{0}^{\infty} \frac{\sum\limits_{i=1}^{n} 1-\CDF{\eig{i}{\mat{X}}}{u}}{1/\rho +u} \D u = \int\limits_{0}^{\infty} \frac{\sum\limits_{i=1}^{n} 1-\CDF{\eig{i}{\mat{Y}}}{u}}{1/\rho +u} \D u \; .
\end{align}
To see the direct part of the Property, recall the Stieltjes transform of a function of bounded variation is in a one-to-one correspondence with the function, and $\sum_{i=1}^{n} 1-\CDF{\eig{i}{\mat{X}}}{u}$ is of bounded variation. It is therefore immediate that if $\E{\LD{\rho \mat{X} }} = \E{\LD{\rho \mat{Y} }}$, then $\sum_{i=1}^{n}\CDF{\eig{i}{\mat{X}}}{u} = \sum_{i=1}^{n}\CDF{\eig{i}{\mat{Y}}}{u}$, a.e.. To prove the converse, assume $\sum_{i=1}^{n} \CDF{\eig{i}{\mat{X}}}{u} = \sum_{i=1}^{n} \CDF{\eig{i}{\mat{Y}}}{u}$ a.e.. Then according to \eqref{J_CapOrd_eqn:app2_M6_1}, $\E{\LD{\rho \mat{X} }} = \E{\LD{\rho \mat{Y} }}$.
\bibliographystyle{IEEEtran}
\nocite{*}
\bibliography{references}

\begin{thebibliography}{10}
\providecommand{\url}[1]{#1}
\csname url@samestyle\endcsname
\providecommand{\newblock}{\relax}
\providecommand{\bibinfo}[2]{#2}
\providecommand{\BIBentrySTDinterwordspacing}{\spaceskip=0pt\relax}
\providecommand{\BIBentryALTinterwordstretchfactor}{4}
\providecommand{\BIBentryALTinterwordspacing}{\spaceskip=\fontdimen2\font plus
\BIBentryALTinterwordstretchfactor\fontdimen3\font minus
  \fontdimen4\font\relax}
\providecommand{\BIBforeignlanguage}[2]{{%
\expandafter\ifx\csname l@#1\endcsname\relax
\typeout{** WARNING: IEEEtran.bst: No hyphenation pattern has been}%
\typeout{** loaded for the language `#1'. Using the pattern for}%
\typeout{** the default language instead.}%
\else
\language=\csname l@#1\endcsname
\fi
#2}}
\providecommand{\BIBdecl}{\relax}
\BIBdecl

\bibitem{paper:AdithyaCapacity12}
A.~Rajan and C.~Tepedelenlioglu, ``Ergodic capacity ordering of fading
  channels,'' in \emph{Proc. IEEE Int. Symp. Inform. Theory}, 2012.

\bibitem{book:verdu}
A.~Tulino and S.~Verd{\'u}, \emph{Random Matrix Theory and Wireless
  Communications}.\hskip 1em plus 0.5em minus 0.4em\relax Now Publishers Inc,
  2004, vol.~1.

\bibitem{paper:letzepis07}
N.~Letzepis and A.~Grant, ``Shannon transform of certain matrix products,'' in
  \emph{Proc. IEEE Int. Symp. Inform. Theory}, 2007, pp. 1646--1650.

\bibitem{quirk62}
J.~Quirk and R.~Saposnik, ``{Admissibility and measurable utility functions},''
  \emph{The Review of Economic Studies}, vol.~29, pp. 140--146, 1962.

\bibitem{belzunce04}
F.~Belzunce and M.~Shaked, ``{Failure profiles of coherent systems},''
  \emph{Naval Research Logistics}, vol.~51, p. 477, 2004.

\bibitem{book:muller02}
A.~M\"{u}ller and D.~Stoyan, \emph{{Comparison Methods for Stochastic Models
  and Risks}}.\hskip 1em plus 0.5em minus 0.4em\relax John Wiley \& Sons Inc,
  2002.

\bibitem{book:shaked}
M.~Shaked and J.~G. Shanthikumar, \emph{{Stochastic Orders and their
  Applications}}, 1st~ed.\hskip 1em plus 0.5em minus 0.4em\relax Springer, Oct.
  1994.

\bibitem{paper:adithya11}
C.~Tepedelenlioglu, A.~Rajan, and Y.~Zhang, ``{Applications of stochastic
  ordering to wireless communications},'' \emph{IEEE Trans. Wireless Commun.},
  vol.~10, pp. 4249 --4257, Dec. 2011.

\bibitem{book:schilling}
R.~Schilling, R.~Song, and Z.~Vondra{\v{c}}ek, \emph{Bernstein Functions:
  Theory and Applications}.\hskip 1em plus 0.5em minus 0.4em\relax Walter de
  Gruyter, 2010.

\bibitem{book:lebedev72}
N.~Lebedev and R.~Silverman, \emph{Special Functions and their
  Applications}.\hskip 1em plus 0.5em minus 0.4em\relax Dover, 1972.

\bibitem{book:bhatia1997}
R.~Bhatia, \emph{Matrix Analysis}.\hskip 1em plus 0.5em minus 0.4em\relax
  Springer Verlag, 1997, vol. 169.

\bibitem{paper:matrixFunctionPaper}
D.~Kressner, ``Bivariate matrix functions,'' in \emph{(online)
  http://sma.epfl.ch/~anchpcommon/publications/multivariate.pdf}, 2012.

\bibitem{book:widder46}
D.~Widder, \emph{{The Laplace Transform}}.\hskip 1em plus 0.5em minus
  0.4em\relax Princeton Univ., 1946.

\bibitem{paper:alouini03}
M.~Hasna and M.-S. Alouini, ``{End-to-end performance of transmission systems
  with relays over Rayleigh-fading channels},'' \emph{IEEE Trans. Wireless
  Commun.}, vol.~2, pp. 1126 -- 1131, Nov. 2003.

\bibitem{paper:waqar10}
O.~Waqar, D.~McLernon, and M.~Ghogho, ``{Exact evaluation of ergodic capacity
  for multihop variable-gain relay networks: A unified framework for
  generalized fading channels},'' \emph{IEEE Trans. Vehicular Technol.},
  vol.~59, pp. 4181 --4187, Oct. 2010.

\bibitem{pun07}
M.~Pun, V.~Koivunen, and H.~Poor, ``Performance analysis of joint opportunistic
  scheduling and receiver design for {MIMO}-{SDMA} downlink systems,''
  \emph{IEEE Trans. Commun.}, pp. 268--280, Jan. 2011.

\bibitem{book:cover}
T.~Cover and J.~Thomas, \emph{{Elements of Information Theory}}.\hskip 1em plus
  0.5em minus 0.4em\relax Wiley-India, 1999.

\bibitem{paper:cioffi01}
W.~Rhee and J.~Cioffi, ``Ergodic capacity of multi-antenna gaussian
  multiple-access channels,'' in \emph{Proc. Thirty-Fifth Asilomar Conf.
  Signals, Systems and Computers}, vol.~1, 2001, pp. 507--512.

\end{thebibliography}
\newpage

\begin{figure}[htb]
\centering
\includegraphics[scale=0.5]{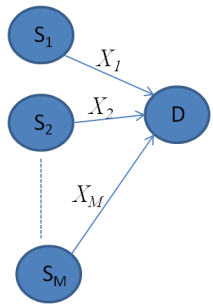}
\caption{$M$-user multiple access channel.}\label{J_CapOrd_fig:mac}
\end{figure}

\end{document}